# Atom- Photon Entanglement in Double- Lambda Quantum System by using femtosecond Gaussian pulses

Zeinab Kordi[1], Seyed Hamideh Kazemi, Saeed Ghanbari and Mohammad Mahmoudi

Department of Physics, University of Zanjan, University Blvd, 45371-38791,

Zanjan, Iran

**Abstract:** The dynamical behavior of entanglement between the atomic ensemble and the spontaneous emission is investigated by applying the femtosecond Gaussian pulses in double-lambda quantum system. In fact by solving the density matrix equations of motion in such a system and by using the von Neumann entropy, the degree of entanglement (DEM) could be measured in semi classical regime and in the multi-photon resonance condition. We interested in investigating the maximum value for DEM and controlled the DEM via relative phase. At first we consider three laser fields as CW optical laser fields and one femtosecond Gaussian field. In this case DEM has interesting behavior. When the system is applied by the other laser fields and Gaussian pulse is still there DEM can be controlled via phase but after finishing the time duration of the femtosecond Gaussian pulse DEM will be phase independent. Then we consider all the laser fields as femtosecond Gaussian pulses which interact with atomic system and we achieve to this point that the steady state entanglement can be occurred by using such ultra-short laser pulses; in this situation DEM will be phase independent. The system would be disentangled in special parameters of Rabi frequencies and all the atoms are coherently populated in a dark state in dressed state picture. Moreover when the system is entangled, there is a population distribution of the atoms in dressed states.

[1] Kordi_z@znu.ac.ir

**Introduction**

In recent years, working and simulating with a computer at high speed in processing the information is a very strong tool to study a lot of phenomena in the nature [1]. In fact there are a lot of physicists and engineers who are working together to achieve a computer performing operations on data according to quantum mechanical phenomena, such as superposition and entanglement [2]. Entanglement had been presented by Einstein, Podolsky and Rosen [3] as an example to show that quantum mechanics could not explain the whole nature based on reality and thus was incomplete. After EPR paper, entanglement has aroused a great deal of interest. The term "Entanglement" has been coined by Schrodinger in his response to Einstein's letter [4] in 1935 on the foundations of quantum mechanics. There were also a lot of philosophical discussions over this bizarre and beautiful aspect of quantum world. So far quantum mechanics is winner in this competition to be consistent with physics of nature.

Mathematically, the entangled state occurs when quantum state of a system consisting of two sub-systems, cannot be described by a simple product of the quantum sates of the two components [5]. In this case, through measurements performed on one of the sub-system, information about the other sub-system can be obtained. Not only does entanglement play an outstanding role in demonstrating Einstein, Podolsky, and Rosen (EPR) paradox [3] and in testing quantum nonlocality [6], but it can also provide a path to novel quantum technologies [7]; indeed, this fundamental property of quantum mechanics can play a leading role in all aspects of quantum information processing tasks, such as quantum super dense coding [8], quantum teleportation [9], quantum algorithm [10] and quantum networking [11].

Photon-based quantum communication [12] and quantum computation [13] are two particularly promising applications, widely explored in the most recent literature. More recently, Duan *et al.* [14] have showed that a hybrid system of light and matter qubits can address the scalability problem of both fields [15, 16]. Generally, entanglement can be happened due to the interaction between different parts of a system consisting of atoms, photons or a mixture of atoms and photons. In the field of the quantum communication, atom-photon entanglement allows to transmit quantum states over arbitrary long distance using the existing telecommunication fiber technologies [17]. It also has obvious advantages for quantum computation, which is typically based on single spins. In fact, optical photons can carry quantum information over long distance and displays almost negligible decoherence; however, they are difficult to store at a fixed location. The reverse is true for atoms. Therefore, by combining the specific advantages of the both atoms and photons, the efficient creation of an entangled pair of particles would be offered.

Up to now, numerous schemes have been used for generating atom-photon entanglement. For instance, Amniat-Talab *et al.* [18] proposed a relatively robust scheme for generation of the maximally entangled states of an atom and a cavity photon, as well as two photons in two spatially separate high-cavities. In another study, Volz *et al.* [19] observed the entanglement between a single trapped atom and a single photon at a wavelength suitable for low-loss communication over large distances. Notably as pointed out by Vedral *et al.* [20], reduced quantum entropy could be used as a good measure to quantify entanglement for bipartite systems. Also, the behavior of atom- photon entanglement has been extensively studied. In 2006, Fang *et al.* [21] investigated entanglement between a Λ-type three-level atomic system and its spontaneous emission field. Recently, it is shown that the atom-photon entanglement can be controlled by relative phase of the applied fields in three- level atomic system in multi- photon resonance condition (MPR) [22, 23]. More recently we have studied dynamical and stationary behavior of atom-photon entanglement in a three-level V-type closed-loop atomic system in the presence of quantum interference due to the spontaneous emission, beyond the two-photon resonance condition [24]. In our previous research we focused on degree entanglement measure (DEM) between the atomic ensemble which it has been dressed by continuous optical laser fields and its spontaneous emission. In this paper we are interested in investigating DEM by using femtosecond Gaussian pulses.

"Recent advances have led to the generation of laser pulses with durations of the order of 1 attosecond. Ultrashort pulses can be used to probe the properties of matter on extremely short time scales [25]". Boyd in chapter 13 of his book titled "nonlinear optics", presented two reasons for emphasizing the importance of studying of ultrashort pulses: "The first reason is that the nature of nonlinear optical interactions is often profoundly modified through the use of ultrashort laser pulses. The second reason is that ultrashort laser pulses tend to possess extremely high peak intensities, because laser pulse energies tend to be established by the energy-storage capabilities of laser gain media, and thus short laser pulses tend to have much higher peak powers than longer pulses." By the generation of femtosecond and attosecond laser pulses (ultrashort pulses), use of few-cycle-pulse laser fields to study of the mechanical effect of light on the atoms and molecules and thereby control its future, is an active field of research in recent years [26-31]. Jiang et al. [27] showed how an optical trap can be manipulated in order to trap the spherical gold nanoparticles with femtosecond pulses. In another study, a simple scheme was proposed to achieve ultrafast and selective population transfer in four-level atoms by utilizing a single nonlinearly chirped femtosecond pulse [28].

It is worth mentioning that optical properties of double-Λ like atomic system which have higher potential applications, are investigated in several theoretical studies [32-38]. In the point of experimental view, the phase-sensitive electromagnetically induced transparency [39] and efficient nonlinear frequency conversion [40] have been studied in this system. The double-lambda schemes have been used also in the past to generate the non-classical states of light [41] and the entangled beams [42-44]. Noting that this model allows for a closed laser-field interaction loop and it is well known that in multi-photon resonance condition, optical properties of the closed-loop atomic system depend on the relative phase of applied fields [45].

In this paper, we investigate the DEM using femtosecond Gaussian laser pulses in a four-level double-lambda type system. It is demonstrated that the DEM could be created by applying Gaussian laser pulses to the system. Also it could be preserved even after finishing such pulses. We show furthermore that the maximal DEM can be achieved by applying one particular Gaussian laser pulse and three CW optical laser fields. In addition, our results show that the DEM could be controlled by the intensities of the applied fields. One of the interesting points here is that if all fields are Gaussian, the behavior of the entanglement is phase-independent, in spite the fact that the system is in closed loop interaction with the pulse. This point lies in the fact that after finishing the time duration of the pulses, the system is not closed anymore and therefore we expect the DEM to be phase-independent. It is worth to mention here that this work is largely motivated by the recent works of the authors [31, 46] where it was shown that a maximum coherent population transfer is achieved for a special value of the relative phase of applied fields. We have shown explicitly that spontaneous emission plays an essential role to create entangled state. By choosing special parameters for Rabi frequencies the system will be disentangled and we could show a link between coherent population trapping and disentanglement by plotting the population of the atoms in dressed state picture.

## Model

Let us consider a four-level double-Λ like atomic system interacting with a train of ultra-short femtosecond Gaussian pulses as shown in Fig. 1. This set contains two metastable lower states $|1\rangle$ and $|2\rangle$ as well as two excited states $|3\rangle$ and $|4\rangle$. The transitions $|1\rangle$-$|3\rangle$, $|2\rangle$-$|3\rangle$, $|1\rangle$-$|4\rangle$ and $|2\rangle$-$|4\rangle$ are excited by four laser fields. The spontaneous emission rates from level $|i\rangle$ ($i \in \{3,4\}$) to the levels $|j\rangle$ ($j \in \{1,2\}$) are denoted by $2\gamma_{ij}$. The laser coupling of the transition $|i\rangle \to |j\rangle$ is characterized by the frequency $\omega_{ij}$ and the wave vector $k_{ij}$. Let us define the electric fields that interact with the pair of energy levels in our atomic system as follows:

$$\mathrm{E}_{ij} = E_{ij}\hat{n}_{ij}e^{-i(\omega_{ij}t-\vec{k}_{ij}\cdot\vec{r}+\varphi_{ij})} + c.c, \qquad (1)$$

where $\hat{n}_{ij}$ and $\phi_{ij}$ are the polarization unit vector and the absolute phase, respectively. We will be especially interested in investigation of the DEM behavior by applying femtosecond Gaussian pulse, which is an electromagnetic pulse whose time duration is of the order of a picosecond or less. Furthermore, we use Gaussian pulses, pulses that have a waveform described by the Gaussian distribution as depicted in Fig. 1. b).

In mathematics point of view, the electromagnetic driving field with Gaussian profile can be explained concisely as follows:

$$\vec{\mathrm{E}}_{ij} = \vec{\mathrm{A}}_{ij}(r,t)Cos(\omega_{ij}t - \vec{k}_{ij}\cdot\vec{r} + \phi_{ij}),$$
$$\vec{\mathrm{A}}_{ij}(r,t) = \mathrm{A}_0\hat{e}_{ij}\exp\left(\frac{r^2}{w^2} - \frac{t^2}{\tau^2}\right)$$

Where $\vec{\mathrm{A}}_{ij}(r,t)$ refers to the space and time dependent envelope and $\mathrm{A}_0, \hat{e}_{ij}, w$ and $\tau$ denote the maximum amplitude of the field, unit polarization vector, beam waist and temporal width of the pulse respectively. In our model, we consider that femtosecond Gaussian pulses have 20 fs time duration.

The semi-classical Hamiltonian in the rotating wave and dipole approximation reads [47].

$$H = \sum_{j=1}^{4}\mathrm{E}_j|j\rangle\langle j| - \sum_{l=3}^{4}\sum_{m=1}^{2}\hbar g_{lm}e^{-i\alpha_{lm}}|l\rangle\langle m| + \mathrm{H}.C, \qquad (2)$$

Here, $g_{lm} = g_{0lm} \exp(-\frac{r^2}{w^2} - \frac{t^2}{\tau^2})$ is the Rabi frequency and $g_{0lm} = (E_{0lm} \hat{n}_{lm} \cdot \vec{d}_{lm})$. Also, $\vec{d}_{lm}$ corresponds to the dipole moments of transitions and energy of the involved states are denoted by $E_j$ ($j \in \{1,...,4\}$). Moreover, the transition frequencies and laser field detuning are shown by $\overline{\omega}_{ij} = (E_i - E_j)/\hbar$ and $\Delta_{ij} = \omega_{ij} - \overline{\omega}_{ij}$, respectively. The exponents are given by $\alpha_{ij} = \omega_{ij} t - \vec{\kappa}_{ij} \vec{r} + \varphi_{ij}$. Next, by changing the reference frame and using the interaction picture, the Hamiltonian can be derived as [38]

$$H = \hbar(\Delta_{32} - \Delta_{31})\tilde{\rho}_{22} - \hbar\Delta_{31}\tilde{\rho}_{33} + \hbar(\Delta_{32} - \Delta_{31} - \Delta_{42})\tilde{\rho}_{44} - \hbar(g_{31}\tilde{\rho}_{31} + g_{32}\tilde{\rho}_{32} + g_{42}\tilde{\rho}_{42} + g_{41}\tilde{\rho}_{41}e^{-i\Phi} + \text{H.C}), \quad (3)$$

Here, $\rho_{ij} = |i\rangle\langle j|$ and $\tilde{\rho}_{ij}$ (i, j $\in \{1,..,4\}$). The so-called relative phase, the multiphoton resonance detuning, wave vector mismatch and initial phase difference are defined as $\Phi = \Delta t - \vec{K}\vec{r} + \varphi_0$, $\Delta = (\Delta_{32} + \Delta_{41}) - (\Delta_{31} + \Delta_{42})$, $\vec{K} = (\vec{\kappa}_{32} + \vec{\kappa}_{41}) - (\vec{\kappa}_{31} + \vec{\kappa}_{42})$, $\varphi_0 = (\varphi_{32} + \varphi_{41}) - (\varphi_{31} + \varphi_{42})$, respectively. The density matrix elements in the rotating frame and rotating wave approximation could be calculated by using Liouville's theorem as follows:

$$\frac{\partial}{\partial t}\rho_{11} = ig_{31}^*\rho_{31} - ig_{31}\rho_{13} + ig_{41}^*\rho_{41}e^{i\Phi} - ig_{41}\rho_{14}e^{-i\Phi} + 2\gamma_{13}\rho_{33} + 2\gamma_{14}\rho_{44},$$

$$\frac{\partial}{\partial t}\rho_{22} = ig_{32}^*\rho_{32} - ig_{32}\rho_{23} + ig_{42}^*\rho_{42} - ig_{42}\rho_{24} + 2\gamma_{23}\rho_{33} + 2\gamma_{24}\rho_{44},$$

$$\frac{\partial}{\partial t}\rho_{33} = -ig_{31}^*\rho_{31} - ig_{32}^*\rho_{32} + ig_{31}\rho_{13} + ig_{32}\rho_{23} - 2\gamma_3\rho_{33},$$

$$\frac{\partial}{\partial t}\rho_{12} = i(\Delta_{32} - \Delta_{31})\rho_{12} + ig_{31}^*\rho_{32} - ig_{32}\rho_{13} + ig_{41}^*\rho_{42}e^{i\Phi} - ig_{42}\rho_{14} - \Gamma_{12}\rho_{12},$$

$$\frac{\partial}{\partial t}\rho_{13} = -i\Delta_{31}\rho_{13} + ig_{31}^*(\rho_{33} - \rho_{11}) - ig_{32}^*\rho_{12} + ig_{41}^*\rho_{43}e^{i\Phi} - \Gamma_{13}\rho_{13},$$

$$\frac{\partial}{\partial t}\rho_{14} = i(\Delta_{32} - \Delta_{31} - \Delta_{42})\rho_{14} + ig_{41}^*e^{i\Phi}(\rho_{44} - \rho_{11}) - ig_{42}^*\rho_{12} + ig_{31}^*\rho_{34} - \Gamma_{14}\rho_{14},$$

$$\frac{\partial}{\partial t}\rho_{23} = -i\Delta_{32}\rho_{23} + ig_{32}^*(\rho_{33} - \rho_{22}) - ig_{31}^*\rho_{21} + ig_{42}^*\rho_{43} - \Gamma_{23}\rho_{23},$$

$$\frac{\partial}{\partial t}\rho_{24} = -i\Delta_{42}\rho_{24} + ig_{42}^*(\rho_{44} - \rho_{22}) - ig_{41}^*\rho_{21}e^{i\Phi} + ig_{32}^*\rho_{34} - \Gamma_{24}\rho_{24},$$

$$\frac{\partial}{\partial t}\rho_{34} = -i(\Delta_{42} - \Delta_{32})\rho_{34} + ig_{31}\rho_{14} + ig_{32}\rho_{24} - ig_{41}^*\rho_{31}e^{i\Phi} - ig_{42}^*\rho_{32} - \Gamma_{34}\rho_{34},$$

$$\rho_{44} = 1 - \rho_{11} - \rho_{22} - \rho_{33}, \tag{4}$$

Where equations are constrained by $\tilde{\rho}_{ij} = \tilde{\rho}_{ji}^*$ and $\sum_i \tilde{\rho}_{ii} = 1$. We define $\gamma_j = \gamma_{1j} + \gamma_{2j}$ and use $\Gamma_{ij} = (2\gamma_i + 2\gamma_j)/2$ ($i \in \{1,2\}$) and ($j \in \{3,4\}$) to denote the damping rate of the coherence on the transition $|i\rangle \to |j\rangle$. For simplicity, the spontaneous emission rates of the excited levels are assumed to be equal. To work out these equations, the phase matching ($\vec{K} = 0$) and multi-photon resonance ($\Delta = 0$) conditions should be satisfied by the applied fields.

In the following, we proceed with the evolution of entropy and measuring DEM. At the moment, measurement of the entanglement is a demanding task and several definitions have been used such as the partial entropy of entanglement [20], the relative entropy of entanglement [5] and entanglement of formation [48]. Here we use the reduced entropy in order to quantify DEM [49]. For a bipartite quantum pure system composed by two subsystems A and B, the partial density operator of one subsystem can be obtained by tracing over the other [50]. The system is called separable, when its density operator can be written as $\rho_{AB} = \rho_A \otimes \rho_B$ with $\rho_{A(B)}$ being the partial density operator of the subsystem A (B), otherwise it is said to be entangled. For this arbitrary bipartite system, the partial von Neumann entropy is defined as [51]

$$S_{A(B)} = -Tr(\rho_{A(B)}) Ln \rho_{A(B)}, \tag{5}$$

Before the interaction between the atomic system and the vacuum fields, the system starts from disentangled pure state which means ($\rho_{33} = 1$) and the excited states are empty of population. Because of the interaction, the reduced entropy of each subsystem will be increased. The entropies of two subsystems are exactly equal at all time and entropy of the one subsystem can be used to measure the DEM.

Phoenix and Knight [52, 53] have shown that decrease in partial entropy indicates that two components evolve toward a pure quantum state. On the contrary, a rise in partial entropy means that they tend to lose their individuality and become correlated or entangled. The DEM for atom-field entanglement is defined as

$$DEM(t) = S_A = S_B = -\sum_{i=1}^{4} (\lambda_i \, Ln \, \lambda_i). \tag{6}$$

Here, $\lambda_i$ is eigenvalue of the partial density matrix.

**Results and discussions**

Now, we are going to depict the results for the behavior of the system in the multi-photon resonance condition. As a realistic example, we consider Rubidium atoms in a vapor cell [44]. In our analyses, all parameters in computer's codes are reduced to dimensionless units through scaling by $\gamma_1 = \gamma_2 = \gamma = 1$ and all plots are sketched in the unit of $\gamma$. We also set $\hbar$ as unity, for convenient.

Firstly, we investigated the time evolution of DEM. We considered an ultra-short femtosecond Gaussian pulse which is applied to level $|1\rangle$ and $|4\rangle$. The other energy levels are applied by the optical laser fields. In Fig.2, the DEM is plotted for different values of relative phase; $\varphi_0 = 0$ (solid), $\pi/2$ (dashed) and $\pi$ (dotted). The other parameters are $\gamma_{13} = \gamma_{23} = \gamma_{14} = \gamma_{24} = \gamma = 1$, $\gamma_3 = \gamma_{13} + \gamma_{23}$, $\Gamma_{13} = \gamma_{13} + \gamma_{23}$, $\Gamma_{23} = \gamma_{13} + \gamma_{23}$, $\Gamma_{24} = \gamma_{14} + \gamma_{24}$, $\Gamma_{14} = \gamma_{14} + \gamma_{24}$, $\Gamma_{34} = \gamma_{13} + \gamma_{23} + \gamma_{14} + \gamma_{24}$, $\Gamma_{12} = \gamma$, $\Delta_{42} = \Delta_{32} = \Delta_{31} = 0$, $g_{32} = 3\gamma$, $g_{31} = 3\gamma$, $g_{42} = 3\gamma$, $r = 50, w = 100, \tau = 20$. As investigation on Fig. 2 shows that we achieve the maximum value of DEM as well as the steady state entanglement. Noting that the time duration of the Gaussian pulse is 20 fs. When the pulse goes in the atomic system, DEM will be increased and the system exhibits an oscillatory behavior; in fact, before that the pulse comes out through the system, DEM is phase dependent so that we could control the entanglement via relative phase. It is worthy of note, however, that after the pulse removal there is still entanglement in the system, it is phase independent. With hindsight, we can say that the atomic system is not closed after finishing the Gaussian pulse and so we expected that the DEM behavior will be phase independent in this open loop configuration.

Then let us move on to Fig. 3, which we consider all the applied fields as femtosecond Gaussian pulses. The other parameters are the same as Fig. 2. In this figure, we have shown a good comparison between the effect of CW laser fields in our previous studying results [46] and the ultra-short femtosecond Gaussian pulses for creating the maximal entanglement. DEM will be decreased and cannot be controlled via relative phase by carefully contrasting in Fig. 3. It is therefore correct to say, that DEM is quite sensitive towards the applied fields and we prefer to consider just one of the applied fields as a Gaussian pulse.

In Fig. 4, the dynamical behavior of DEM is plotted for different values of time duration parameters, i.e. $\tau = 20\,fs$ (solid), $\tau = 40\,fs$ (dashed) and $\tau = 60\,fs$ (dotted). The relative phase is

$\varphi_0 = \pi$ and the other parameters are same as Fig. 2. It is found out that the oscillatory behavior of DEM is related to the scattering of the CW laser fields into the ultra-short femtosecond Gaussian beam mode, at a different frequency of this beam. When the Gaussian beam is finished, the scattering is omitted and DEM shows the steady state behavior.

Below we would like to draw attention to the fact that the spontaneous emission, which has this potential of changing the population of the atomic excited levels, is a good source to create entangled system [54,55].

Fig. 5 shows the dynamical behavior of DEM for different values of intensity of optical fields; $g_{41} = 0.03\gamma$, $g_{42} = 3\gamma$ (a) and $g_{41} = g_{42} = 3\gamma$ (b).

For understanding the physics of entanglement between the atom and the spontaneous emission, we investigate the population of the atoms among the dressed states. Note that the system will be dressed and the atoms could be found in dressed states, when the atomic system is applied by some optical laser fields. In order to emphasize on the effect of spontaneous emission in this study, let us turn off the optical laser fields which coupled the transitions $|1\rangle - |3\rangle$ and $|2\rangle - |3\rangle$. In the initial time, the only state $|3\rangle$ is populated and the other levels are empty of the populations.

The calculated dressed states in the absence of two of the optical driving fields, i.e., $g_{31} = g_{32} = 0$, as follows:

$$|A\rangle = |3\rangle,$$

$$|B\rangle = -\frac{g_{42}}{\sqrt{g_{42}^2 + g_{41}^2}}|1\rangle + \frac{g_{41}}{\sqrt{g_{42}^2 + g_{41}^2}}|2\rangle,$$

$$|C\rangle = \frac{g_{41}}{\sqrt{2(g_{41}^2 + g_{42}^2)}}|1\rangle + \frac{g_{42}}{\sqrt{2(g_{41}^2 + g_{42}^2)}}|2\rangle + \frac{1}{\sqrt{2}}|4\rangle,$$

$$|D\rangle = -\frac{g_{41}}{\sqrt{2(g_{41}^2 + g_{42}^2)}}|1\rangle - \frac{g_{42}}{\sqrt{2(g_{41}^2 + g_{42}^2)}}|2\rangle + \frac{1}{\sqrt{2}}|4\rangle,$$

The dynamical behaviors of populations of different dressed states atoms are plotted In Fig. 6. Used parameters are $g_{41}=0.03\gamma$, $g_{42}=3\gamma$ (a) and $g_{41}=g_{42}=3\gamma$ (b). Other parameters are the same as in Fig. 2. As it can be observed from Fig. 6 (a), whole population is transferred from the upper level $|3\rangle$ to the dressed state $|B\rangle$ which is superposition of the energy level $|1\rangle$ and energy level $|2\rangle$, showing the creation of atomic coherence by spontaneous emissions [54]. At first one the system is disentangled because the coupling field of the transition $|1\rangle-|4\rangle$ is not enough strong to pump the atoms, which they are transferred of level $|3\rangle$ to level $|1\rangle$ by spontaneous emission, from the ground state to excited state. As we know the state $|3\rangle$ is not dressed with any optical laser field and its dressed state is equal to its bare state.

Let us increase the Rabi frequency according to Fig. 6 (b) ($g_{41}=3.0\gamma$). By increasing the intensity of this applied field the atomic system will be entangled with its spontaneous emission. We have shown that DEM could be controlled via intensity of applied fields in the presence of spontaneous emission.

**Acknowledgments**

**Figures captions**

**Figure 1.** Schematic diagram of the closed-loop four-level double-Λ quantum system. This system is driven by optical laser fields and train of ultra-short femtosecond Gaussian pulses Fig.1. b). Spontaneous decays are denoted by the wiggly green lines.

**Figure 2.** Time evolution of DEM in closed-loop configuration for $\varphi_0 = 0$ (solid), $\pi/2$ (dashed) and $\pi$ (dotted). The parameters are $\gamma_{13} = \gamma_{23} = \gamma_{14} = \gamma_{24} = \gamma = 1$, $\gamma_3 = \gamma_{13} + \gamma_{23}$, $\Gamma_{13} = \gamma_{13} + \gamma_{23}$, $\Gamma_{23} = \gamma_{13} + \gamma_{23}$, $\Gamma_{24} = \gamma_{14} + \gamma_{24}$, $\Gamma_{14} = \gamma_{14} + \gamma_{24}$, $\Gamma_{34} = \gamma_{13} + \gamma_{23} + \gamma_{14} + \gamma_{24}$, $\Gamma_{12} = \gamma$, $\Delta_{42} = \Delta_{32} = \Delta_{31} = 0$, $r = 50, w = 100, \tau = 20$, $g_{31} = 3\gamma, g_{42} = 3\gamma$, $g_{32} = 3\gamma$ and $g_{41}$ is considered as a femtosecond Gaussian pulse.

**Figure 3.** Time evolution of DEM in closed-loop configuration. In this figure all the fields are femtosecond Gaussian pulse. Other parameters are same as Fig. 2.

**Figure 4.** Dynamical behavior of DEM for different values of time duration $\tau = 20\ fs$ (solid), $\tau = 40\ fs$ (dashed) and $\tau = 60\ fs$ (dotted). The relative phase is $\varphi_0 = \pi$ and the other parameters are same as Fig. 2.

**Figure 5.** DEM behavior versus time for $g_{41} = 0.03\gamma$, $g_{42} = 3\gamma$ (a) and $g_{41} = g_{42} = 3\gamma$ (b).

**Figure 6.** dynamical behavior of the population in dressed-state picture for parameters $g_{41} = 0.03\gamma$, $g_{42} = 3\gamma$ (a) and $g_{41} = g_{42} = 3\gamma$ (b).

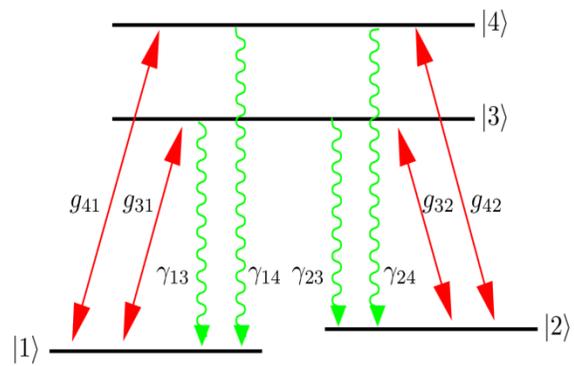

Fig. 1.a)

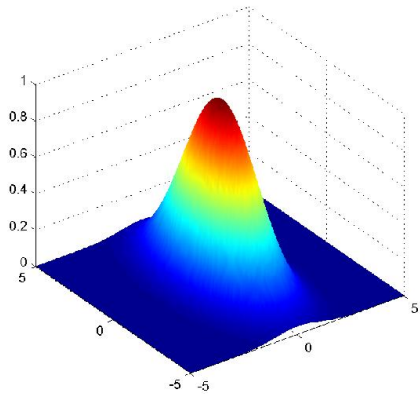

Fig. 1.b)

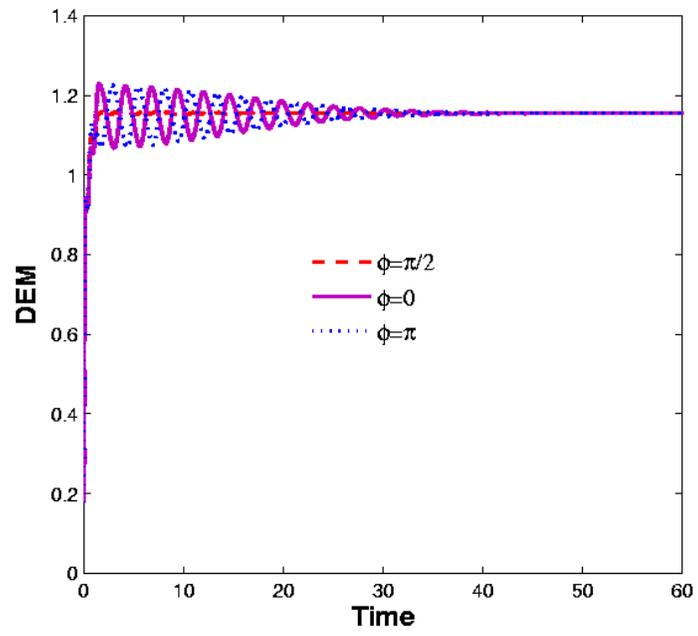

Fig. 2.

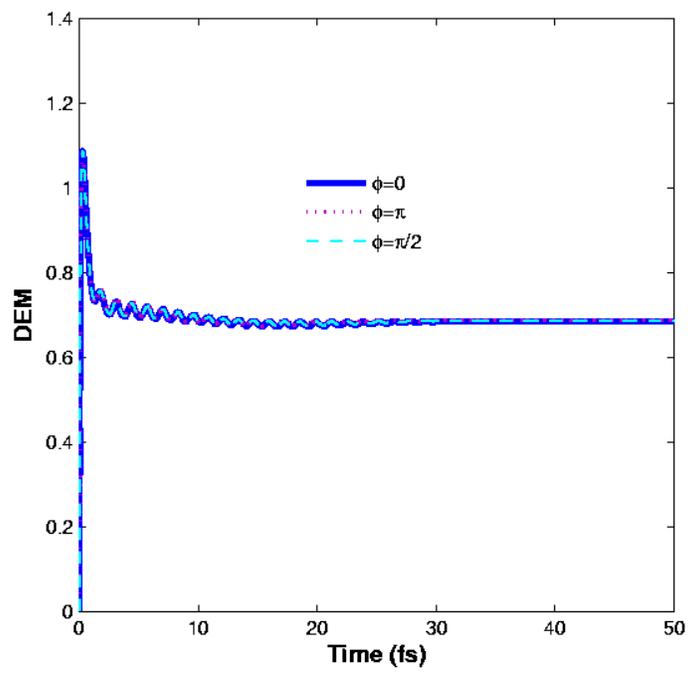

Fig. 3.

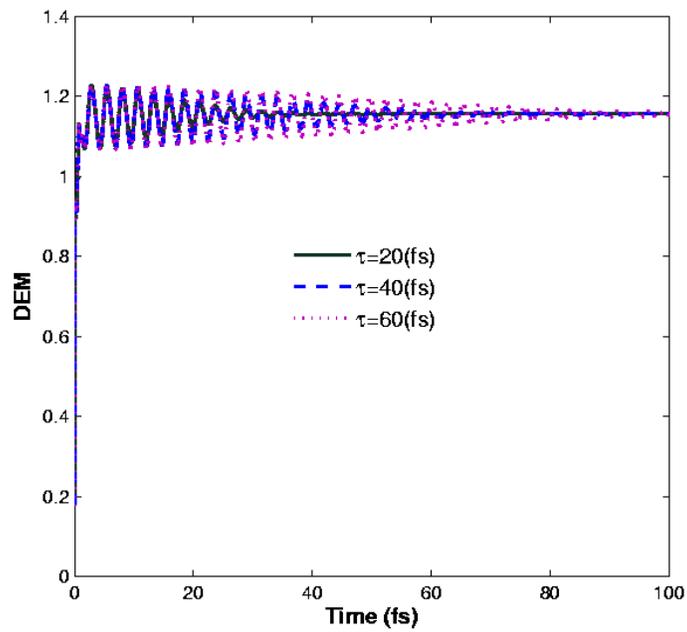

Fig. 4.

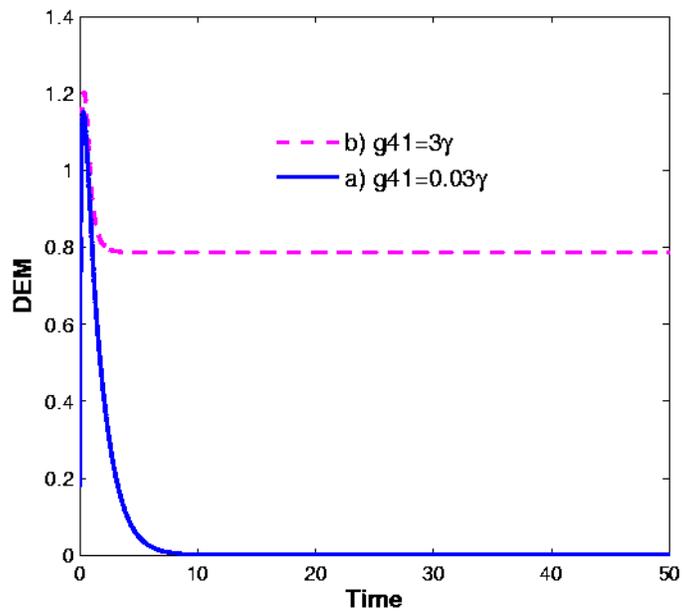

Fig. 5.

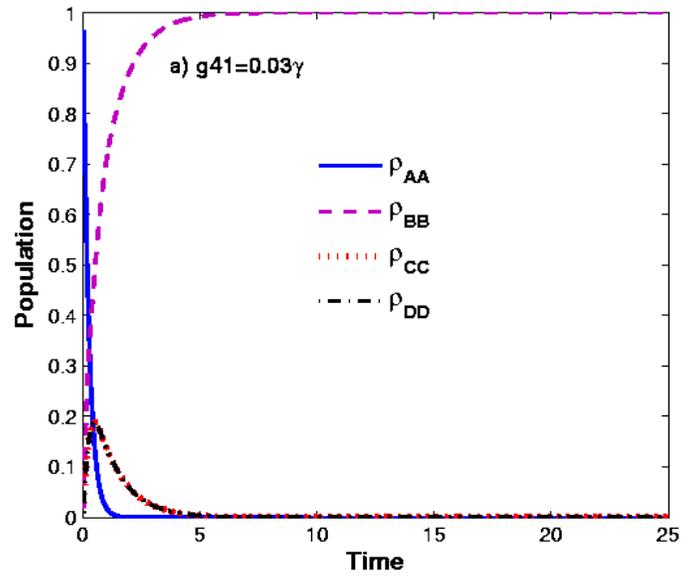

Fig. 6. a)

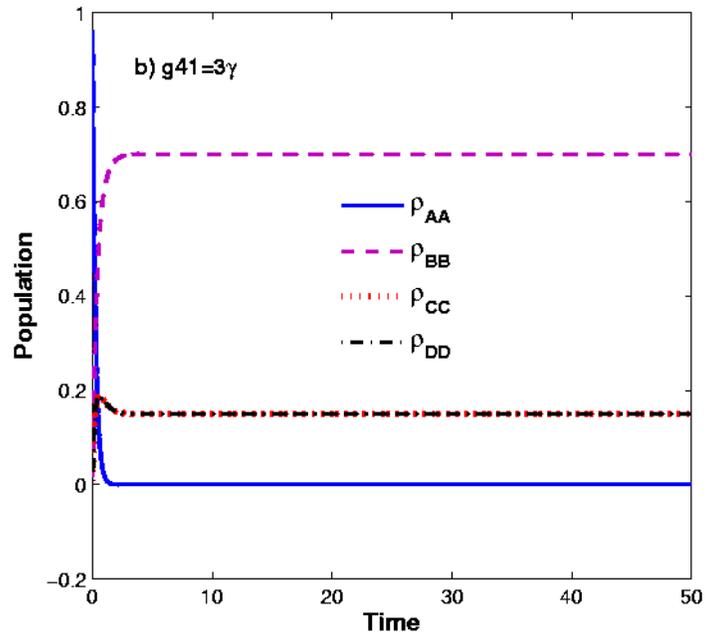

Fig. 6. b)